# Organic Crystal Active Waveguide as an All-Angle Signal Receiver and Transmission Platform for Visible Light Communication

*Ankur Khapre,* [#] *Jyotisman Hazarika,* [#] *Rajadurai Chandrasekar**

A. Khapre,[#] J. Hazarika,[#] Prof. R. Chandrasekar

School of Chemistry and Centre for Nanotechnology, University of Hyderabad,

Prof. C. R. Rao Road, Gachibowli, Hyderabad 500046, India

E-mail: r.chandrasekar@uohyd.ac.in

[#] Equal contribution of authors

Abstract

Organic crystal waveguides, known for their excellent light-guiding and photonic versatility, offer a promising alternative to conventional optical media in visible light communication (VLC) systems. In an unprecedented approach, a robust and high photoluminescence quantum yield organic crystal, 2,2'-((1*E*,1'*E*)-hydrazine-1,2-diylidenebis(methaneylylidene))diphenol (SAA), is used as an optical waveguide medium for the real-time data communication in the visible range utilizing a microcontroller unit based on–off keying modulation. Employing the spectral properties, the versatility of the organic crystal to modulate signals *via* dual active and passive waveguiding is demonstrated. An error-free signal detection is achieved due to the smooth, defect-free surface morphology of the crystal. The incident angle versus light intensity reveals stable fluorescence under narrow-angle excitation, highlighting the crystal's unique capability as an all-angle signal receiver, surpassing conventional optical fibers. A prototypical real-time data transfer setup is realized, capable of transmitting streams directly from the serial interface and reconstructing grayscale images with high fidelity. Overall, we report the first implementation of an all-organic crystal-based VLC platform that integrates wavelength conversion, all-angle waveguiding, and real-time signal processing, laying the foundation for compact, efficient, and integrable photonic communication technologies.

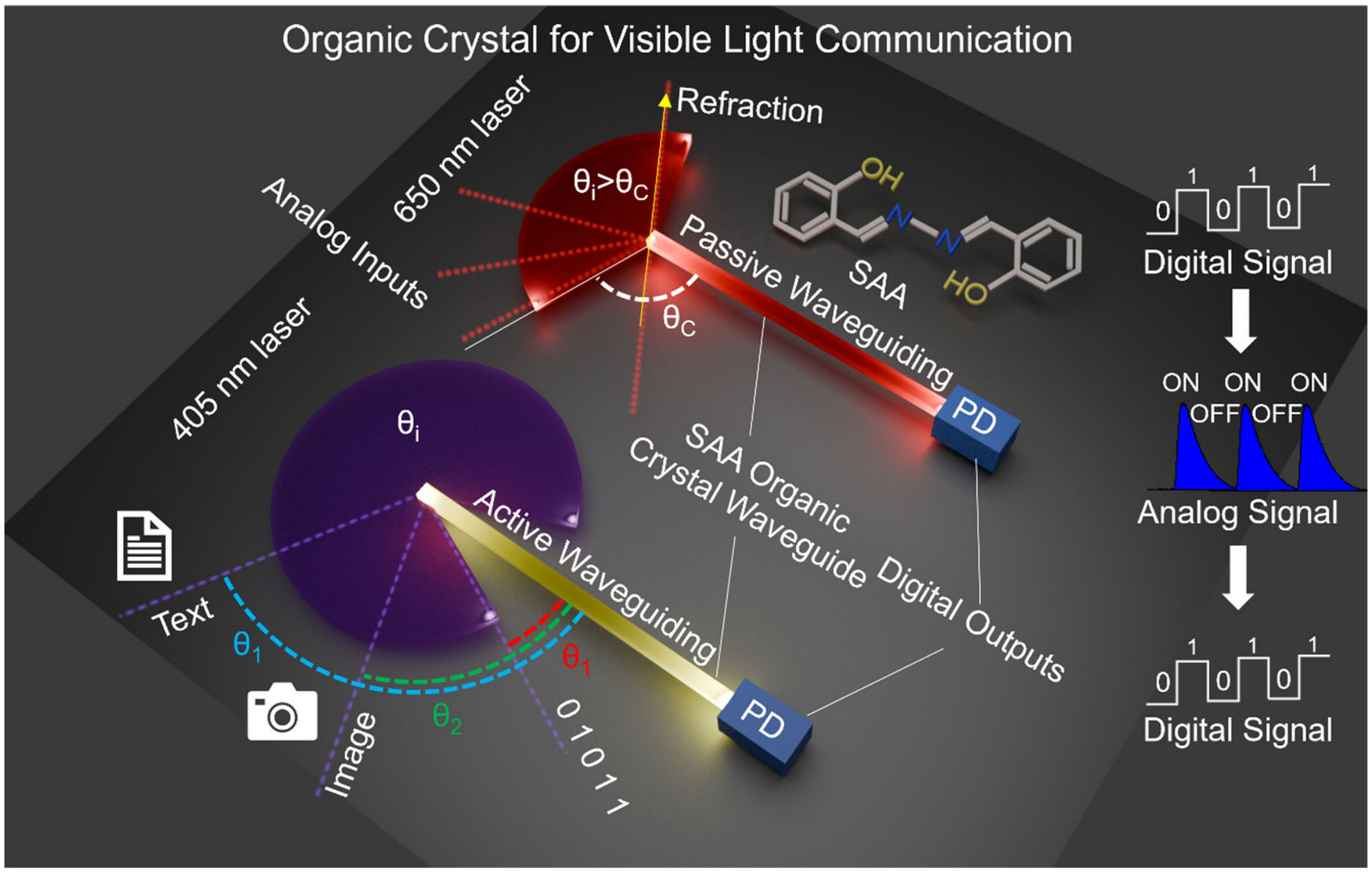

**Scheme 1.** Graphical representation of dual-mode waveguiding in an SAA organic single crystal. Under narrow-angle excitation ($\theta_i < \theta_c$), only active waveguiding *via* 405 nm laser yields signal output, while passive waveguiding with 650 nm laser fails to transmit data, highlighting the angular tolerance unique to active VLC operation.

## 1. Introduction

Visible light communication (VLC) has emerged as a promising alternative to traditional radio frequency communication, particularly in the context of growing demands for high-speed, energy-efficient, and secure data transmission systems [1]. Exploiting the visible light of the electromagnetic spectrum (400–700 nm), VLC enables simultaneous illumination and communication, offering the advantages of unregulated bandwidth, minimal electromagnetic interference, and enhanced physical security owing to its limited penetration through walls [2]. These attributes have rendered VLC a compelling candidate for applications ranging from indoor wireless networking to underwater communications and vehicular systems [3].

A key technological requirement for efficient VLC systems is the development of robust optical materials capable of modulating, guiding, and converting light with high fidelity. Traditionally, inorganic semiconductors and silica-based optical fibers have dominated the field, benefiting from their well-understood optical properties and mechanical robustness [4].

However, optical fibers primarily function as passive waveguides, relying solely on total internal reflection and offering limited functionality in terms of wavelength conversion or active light manipulation [5]. Moreover, their rigid structure and dependence on precise fabrication methods impose limitations for miniaturized, flexible, and integrated photonic applications. In contrast, organic materials, particularly organic crystals, have attracted increasing interest for their potential as multifunctional optical elements in VLC and photonic systems [6]. Organic single crystals exhibit a combination of optical anisotropy, high photoluminescence quantum yields (PLQY), low optical losses, and tunable absorption-emission characteristics, attributes that are highly desirable for flexible optoelectronics. The flexibility in molecular design further allows tailoring their optical bandgaps to specific communication wavelengths, providing advantages over traditional inorganic platforms [7].

Among various photonic functionalities in organic crystals, the ability to perform active waveguiding, where absorbed light is emitted *via* fluorescence (FL) and then self-guided along the crystal, represents a significant advancement [8]. Active waveguiding enables wavelength conversion and amplification processes that are difficult to achieve with conventional optical fibers. Meanwhile, passive waveguiding in organic crystals can be achieved through refractive index contrast, similar to fiber optics, with the added advantage of broader material tunability and easier integration onto diverse substrates [9]. Importantly, the angular dependence of optical signal transmission, which imposes constraints on fiber-based systems, can be significantly relaxed in organic active waveguides due to the nature of their absorption and subsequent waveguiding mechanisms. Recent efforts have explored the integration of organic materials into VLC systems. Organic photodetectors and light-emitting diodes have been proposed for VLC components, offering low-cost and scalable solutions [10]. However, the exploration of organic crystals as simultaneous waveguides and wavelength converters within VLC setups remains relatively underexplored, despite their promising material characteristics.

In this study, we employ an organic single crystal of 2,2'-((1*E*,1'*E*)-hydrazine-1,2-diylidenebis(methaneylylidene))diphenol (SAA) as optical receiver and transmission medium within the VLC framework. SAA possesses a broad absorption band extending up to approximately 500 nm and a strong FL in the visible range between 495 and 720 nm [11], making it an ideal candidate for supporting both active and passive optical waveguiding processes (**Scheme 1**). By carefully choosing the excitation wavelength relative to the material's absorption-emission profile, we demonstrate two distinct regimes: active waveguiding under violet (405 nm) excitation, where the absorbed light emits as FL and guides, and passive waveguiding under red (650 nm) excitation, where the light is transmitted without absorption. The crystal's stable performance across varied excitation angles and dual waveguiding functionality enables robust signal propagation without the need for external wavelength-selective components. Additionally, we implement a complete VLC architecture consisting of a binary light-pulse encoding transmitter using on–off keying (OOK) modulation and a visible-range photodetector-based receiver capable of signal reconstruction [12]. Through this work, we demonstrate that organic crystals can serve as compact, multifunctional, and efficient platforms for VLC, offering wavelength conversion, angular tolerance, and transmission reliability beyond what is achievable with conventional optical media, thereby laying the foundation for flexible, next-generation optical communication systems.

## 2. Results and discussion

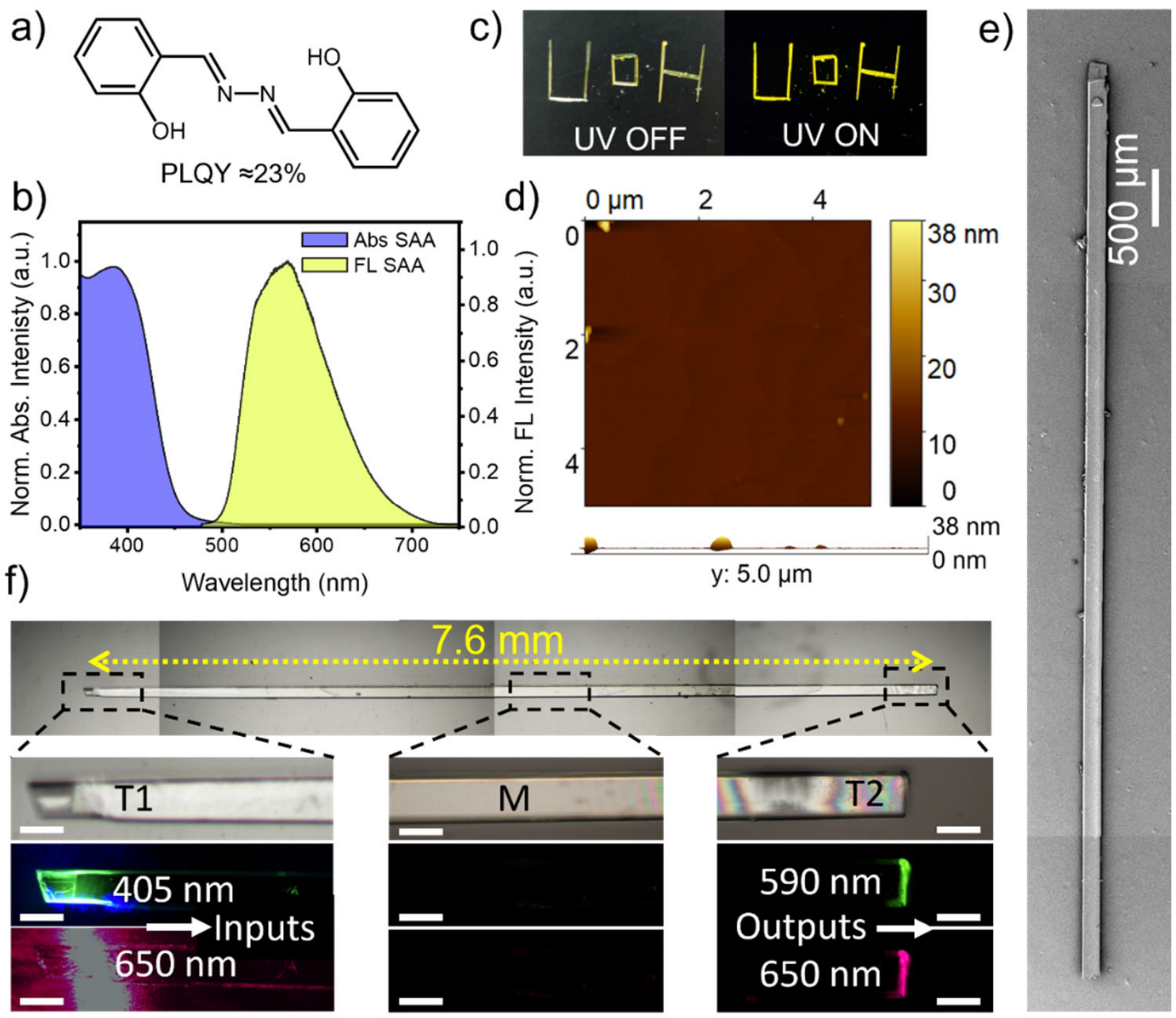

**Figure 1.** a) Molecular structure of SAA. b) The room temperature solid-state absorption and emission spectra of SAA and its c) photographs in ambient conditions and under UV light. d) Top and side view AFM topography and e) FESEM image of a typical SAA single crystal. f) Optical images of a 7.6 mm SAA crystal showing its efficient light-guiding ability. Insets show the bright field and FL images for active and passive waveguiding. The scale bar corresponds to 100 μm.

Single crystals of SAA were grown by slow evaporation from a mixture of ethanol and dichloromethane (**Figure 1**a). The solid-state absorption-emission spectra of SAA crystals revealed optical absorption in the UV–visible region up to 500 nm and FL in the yellow region ($\lambda \approx$ 495-720 nm; $\lambda_{max} \approx$ 568 nm) of the visible spectrum (Figure 1b, c). The SAA crystals, previously reported to have excellent mechanical properties [6a], were devoid of physical defects. Atomic force microscope (AFM) topography and field emission scanning electron microscope (FESEM) imaging of a regular SAA crystal confirmed the smooth and well-defined surface morphology of the crystals (Figure 1d, e). To evaluate the waveguiding capability of SAA crystals, a crystal of approximately 7.6 mm in length was selected and examined under an optical microscope (Figure 1f). Upon excitation at one tip of the crystal with a 405 nm laser, a bright FL signal was observed at the opposite tip. Similarly, irradiation with a 650 nm laser produced a distinct transmitted signal at the distal tip of the crystal. These observations collectively validated the light-guiding ability of the SAA crystal.

The transmitter system facilitated the optical transmission of a predefined text, "Visible light communication using organic crystal", by implementing OOK modulation. The encoded data was derived from the ASCII representation of each character in the string, where each character was converted into its 8-bit binary equivalent. The transmission followed a least significant bit (LSB) first approach, ensuring sequential encoding of the binary data. A monochromatic light source served as the optical carrier, modulated at a controlled frequency to correspond to the binary states. A logical ‘1’ was represented by the activation of the light source, while a logical ‘0’ corresponded to its deactivation. To ensure accurate transmission, each bit was assigned a fixed pulse width, allowing precise differentiation between high and low states. The modulated optical signal was directed onto an SAA crystal, functioning as an optical channel.

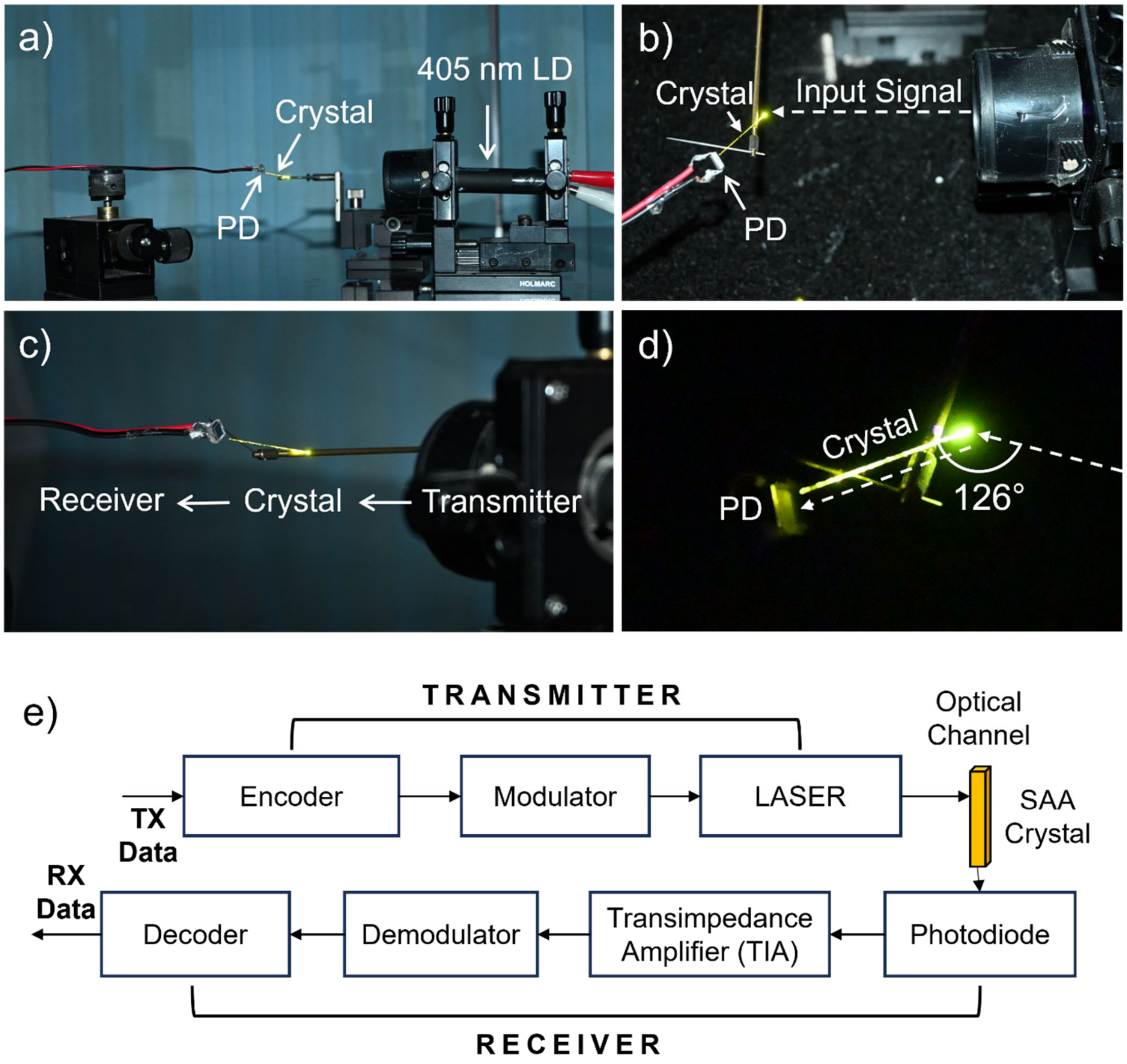

**Figure 2.** a–d) Sequentially zoomed-in photographs of the VLC setup with organic crystal (SAA) waveguide-based optical channel and its e) block diagram. LD and PD stand for Laser diode and Photo diode, respectively.

A nearly 3 cm long SAA crystal was mounted on a probe arm attached to a micropositioner to enable precise alignment within the optical path (**Figure 2**). The micropositioner provided fine control over the crystal's position in the x, y, and z directions, ensuring optimal placement for efficient light propagation and detection. This setup allowed for accurate adjustments to maximize the intensity of the emitted visible light reaching the detector while minimizing alignment errors. Additionally, the controlled positioning facilitated systematic measurements of transmission efficiency and signal stability by varying the crystal's distance and orientation relative to the receiver.

The receiver system detected and decoded the signal transmitted by the organic crystal. A photodiode, chosen for its high sensitivity in the visible spectrum, converted the incident light into a photocurrent, which was amplified using a transimpedance amplifier (TIA). The conditioned signal was digitized by a microcontroller, which sampled intensity variations and deciphered binary data using OOK modulation. A thresholding algorithm reconstructed the 8-bit ASCII characters, restoring the transmitted text.

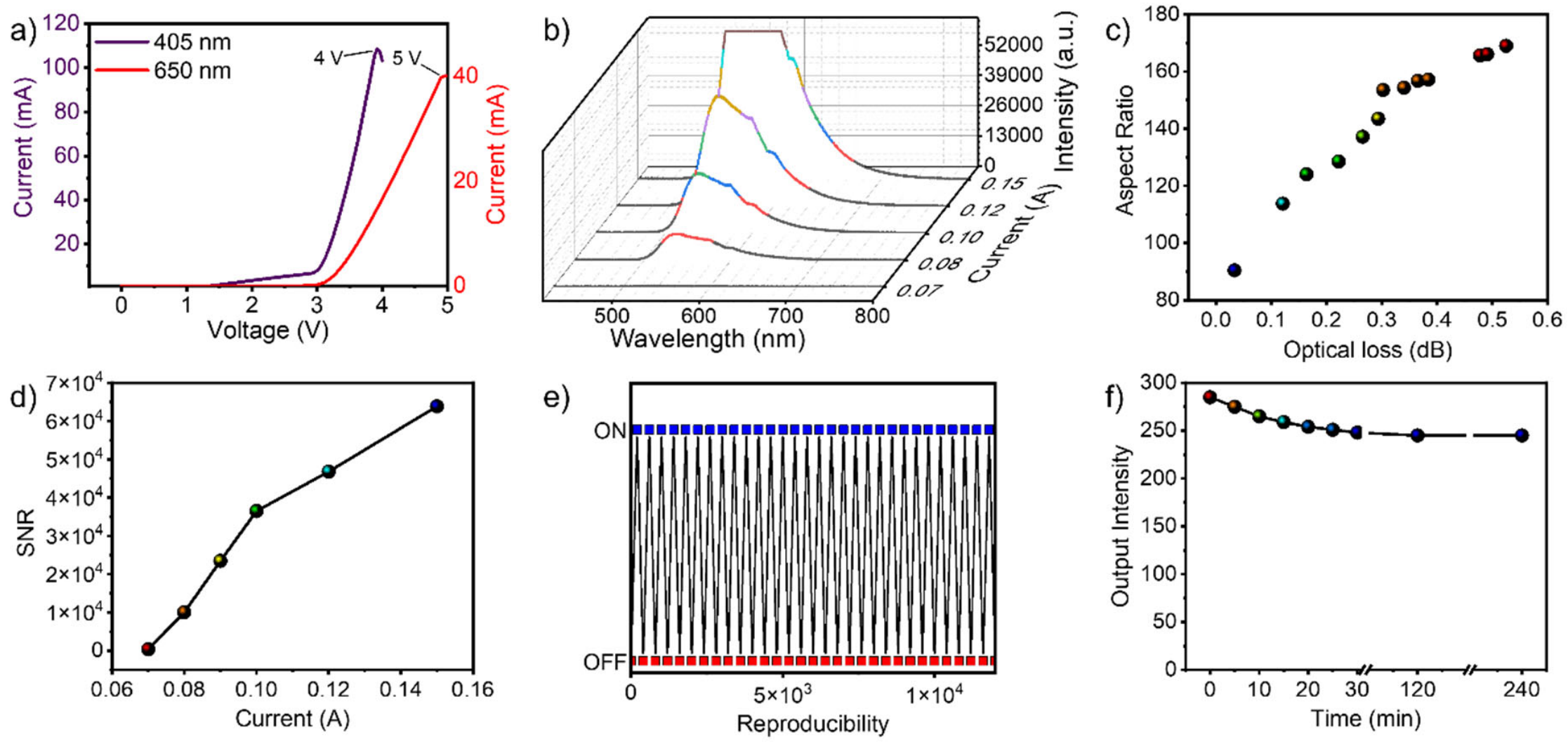

**Figure 3.** a) Current–voltage (I–V) characteristics of 405 and 650 nm laser diodes (LDs) used for excitation in the VLC setup, showing threshold behaviour and current output differences. b) 3D plot showing the variation of FL wavelength and intensity from the SAA crystal as a function of input laser drive current. c) Optical loss in the SAA crystal waveguide as a function of the crystal aspect ratios. d) Signal-to-noise ratio (SNR) as a function of laser drive current, showing a steady increase in SNR with higher excitation power. e) Reproducibility test showing consistent detection of binary pulses through the SAA crystal and its f) temporal stability of light transmission under continuous laser excitation.

Initially, to characterize the electrical behavior of the excitation sources employed in the VLC system, current–voltage (I–V) measurements were carried out for both the 405 and 650 nm laser diodes (LDs). The 405 nm diode (violet line) showed a sharp increase in current beyond a threshold voltage of approximately 3 V, reaching ~100 mA at 4 V (**Figure 3**a). In contrast, the 650 nm diode (red line) reached a maximum current of ~40 mA at 5 V. The steeper I–V slope for the 405 nm diode indicated a higher drive efficiency and current injection rate compared to the 650 nm diode. To investigate the light intensity at the SAA crystal output as a function of the LD drive current, the input current was modulated, ranging from 0.07 A to 0.15 A for a 405 nm diode. The laser power increased with the increment in input current, and a corresponding rise in FL intensity was observed, indicating efficient absorption and emission by the SAA waveguide (Figure 3b). The emission was centered within the expected FL range, with spectral stability maintained across all current values. Notably, the FL intensity reached a saturation threshold near 0.15 A, beyond which no further significant enhancement was recorded. These characteristics were critical for tuning the excitation conditions in the VLC platform, enabling controlled light modulation from the SAA crystal under varied wavelength and power regimes.

For the understanding of how geometrical parameters influence signal attenuation, we investigated the relationship between optical loss and the aspect ratio of the crystal (defined as the length-to-thickness ratio). A clear increasing trend in optical loss was observed with rising aspect ratio, indicating that longer and thinner crystals tend to accumulate higher propagation losses (Figure 3c; Figures S4, S5, Supporting Information). This was attributed to increased surface scattering and reduced optical mode confinement in crystals with reduced thickness and extended length. Further, the signal-to-noise ratio (SNR) was analyzed as a function of the drive current applied to the excitation laser diode (Figure 3d). The SNR exhibited a consistent and approximately linear increase with increasing current, spanning from 0.07 A to 0.15 A. This trend was indicative of enhanced light signal strength from the SAA crystal under higher excitation power, which improved the detectability of the modulated signal against background noise. At the upper limit of 0.15 A, the system achieved a peak SNR exceeding $6 \times 10^4$, ensuring high signal clarity and low bit error rates during data transmission.

To assess the performance and stability of the organic crystal-based VLC system, several other key optical and communication metrics were systematically evaluated. The reproducibility of data transmission was first examined by repeatedly transmitting a predefined binary sequence through the organic crystal waveguide. The high fidelity between the transmitted and received binary signals across more than $10^4$ cycles confirmed the ability of the waveguide to maintain consistent signal integrity without degradation or distortion over time (Figure 3e). The operational stability of the organic crystal waveguide was monitored under continuous illumination to evaluate its photostability and resistance to photobleaching or thermal effects. The output light intensity exhibited a minor decline within the initial 20 minutes of laser excitation, likely due to initial thermal equilibration or minor photochemical changes (Figure 3f). However, after this transient period, the output stabilized and maintained a steady intensity for the remainder of the 4-hour test duration, highlighting the robustness of the crystal for sustained optical transmission applications.

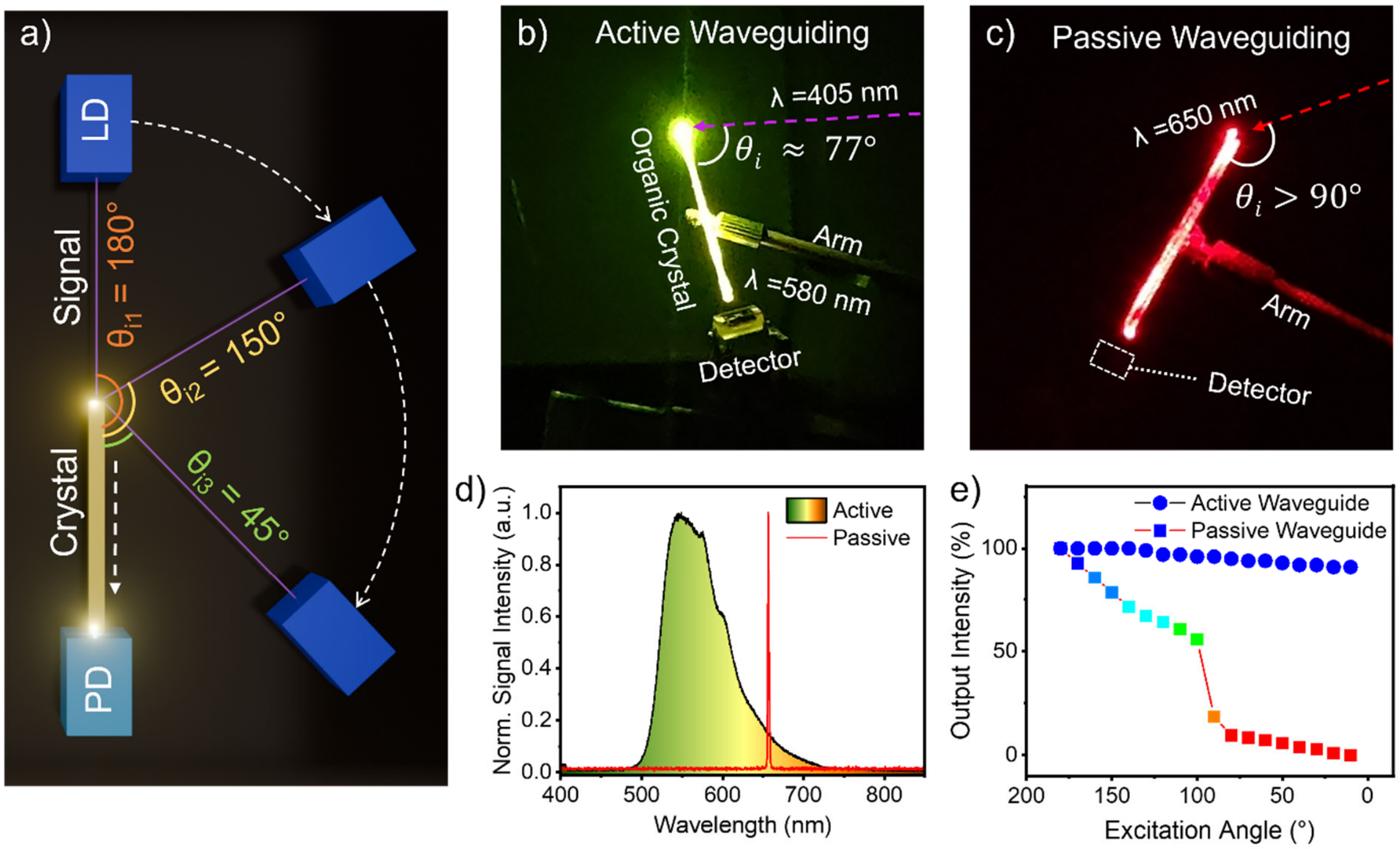

**Figure 4.** a) Schematic illustration for an angle-dependent organic optical channel (LD: Laser Diode, PD: Photodiode). Photographs of SAA crystal mounted on the micropositioner for b) active and c) passive waveguiding. d) The FL spectra recorded at the SAA crystal output for excitations with 405 nm and 650 nm lasers, and their e) plot showing the effect of incident light angle on the FL output for active and passive waveguiding mechanisms, respectively.

After successful characterizations, a 405 nm laser was employed as the excitation source, selected to match the absorption region of the SAA crystal. Upon pulsed with the encoded information, SAA efficiently absorbed the input pulses, undergoing electronic excitation followed by FL in the visible spectrum (**Figure 4**a-b). A spectrometer was used to record the optical spectrum at the crystal output, and it revealed an FL in the region 495–720 nm (Figure 4d). This process exemplified active waveguiding, where the crystal not only facilitated the information through light propagation but also served as a wavelength converter by transferring the information to the generated FL. The guided FL traveled through the crystal structure, enhancing directional emission toward the receiver. The effect of incident light angle on the FL output was systematically investigated by varying the angle of the 405 nm laser excitation from 10° to 180° with an increment of 10° per recording and measuring the corresponding FL transmitted signal from the SAA crystal (Figure S2, Supporting Information). Remarkably, the study revealed that, unlike conventional optical fibers, which impose strict limitations on light propagation due to critical angle constraints, the SAA crystal exhibited relatively stable FL intensity across different incident angles, indicating the highly adaptable active waveguiding behavior (Figure 4e). Even at small incident angles, the 405 nm laser excitation efficiently induced FL, allowing for robust optical signal communication.

In contrast, a 650 nm laser with encoded information as an excitation source propagated through the crystal without wavelength transformation (Figure 4c). The mechanism relied on passive waveguiding, allowing the incident light, which lies outside the absorption region of SAA, to undergo total internal reflection. The FL spectrum recorded at the crystal output showed a sharp peak at 650 nm, which confirmed the non-absorptive transmission of the incident light (Figure 4d). The intensity of the transmitted 650 nm light varied with changes in the incident angle from 10° to 180°, highlighting the dependence of waveguiding efficiency on optical alignment, unlike in active waveguides (Figure S3, Supporting Information). At optimal incident angles ($\theta_i \geq 90°$), the laser light is efficiently coupled into the crystal, propagating through internal reflections. However, at non-ideal angles ($\theta_i \leq 90°$), increased scattering and partial reflection at the crystal-air interface led to a reduction in transmitted light intensity (Figure 4e). This variation underscored the importance of precise angular control in maximizing waveguiding efficiency for efficient optical data communication.

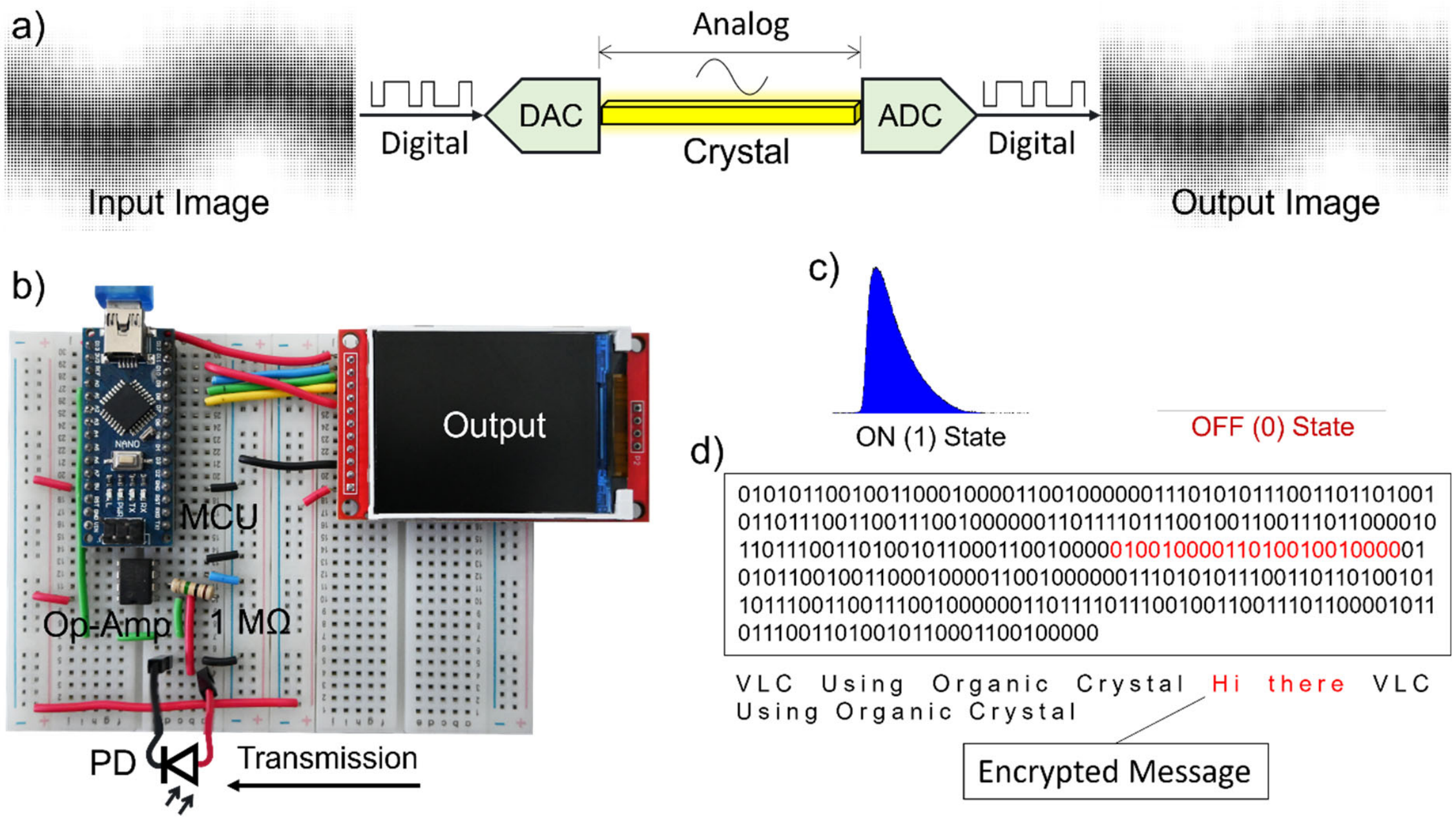

**Figure 5.** a) Schematic illustration showing an image transmission through an SAA organic crystal. b) Physical setup of the VLC receiver circuit. c) Analog signal intensity distribution for ON (1) and OFF (0) states of the SAA organic crystal. d) Encrypted data sequence embedding a visible "Hi there" message within a continuous stream of ASCII binary data, demonstrating proof-of-concept for secure or selective message decoding in the presence of a constant transmission background.

To showcase the practical capabilities of the organic crystal-based VLC system to handle complex data types such as images, a program was implemented for a static image transfer (Figure 5a). A grayscale image was first converted into a binary stream through a digital-to-analog conversion (DAC) process. Each pixel's intensity was mapped to a binary intensity-modulated sequence, which was then transmitted as optical pulses through the organic crystal waveguide. On the receiving end, the photodiode captured the modulated light, which was amplified and processed by an analog-to-digital converter (ADC) to reconstruct the output image. Despite inherent signal attenuation and environmental noise, the recovered image closely resembled the input, validating the waveguide's ability to preserve spatial and intensity information over a large binary data stream.

Finally, we extended the VLC setup to demonstrate real-time serial communication by constructing a live communication prototype (Figure 5b). A transmitter core was used to encode user input from the serial monitor into ASCII binary sequences via OOK modulation. These sequences were optically transmitted through the SAA crystal waveguide and decoded by a PD-based receiver. The reconstructed data was displayed on an LCD in real time, enabling uninterrupted streaming of text and image content. To illustrate its potential in secure communication, a hidden message "Hi there" was embedded within a continuous binary stream containing repeating standard text. This message, distinguishable only through specific parsing of the incoming data, highlights the system's capacity to handle covert transmission scenarios. The successful and distortion-free reconstruction of the embedded message validated the fidelity and robustness of the crystal-based VLC channel, emphasizing its relevance in both conventional and secure optical communication frameworks.

## 3. Conclusion

This study unveiled the potential of an organic crystal, SAA, as a dynamic optical medium for VLC, seamlessly integrating both active and passive waveguiding mechanisms. A transmitter system encoded data using OOK modulation, converting text into binary pulses of monochromatic light sources. The receiver, equipped with a high-sensitivity photodiode, detected the modulated PL and reconstructed the transmitted information. Unlike conventional optical channels such as fibers, the SAA crystal sustained light propagation even under narrow-angle excitation, establishing its capability as an all-angle signal receiver and transmission platform. The crystal exhibited stability in active waveguiding and controlled intensity variations in passive waveguiding, offering an adaptable platform for optical signal transport. By eliminating the need for external wavelength-selective components, it presented a simplified yet powerful alternative for VLC applications. The ability to transmit complex data types such as images, alongside real-time encoded messages, validated the system as a promising platform for secure, compact, and integrable optical communication technologies based on organic materials, where the very air we breathe is infused with invisible signals of light, linking devices and people in a seamless, connected world.

## Supporting Information

Supporting information is available.

## Acknowledgments

AK and JH contributed equally to this work. RC acknowledges SERB-New Delhi [SERB-STR/2022/00011 and CRG/2023/003911] for financial support. AK thanks PMRF for the fellowship.

Supporting Information

**Table of contents:**

## 1. Materials and Electronic components

All chemicals and solvents, salicylaldehyde, hydrazine monohydrate, potassium carbonate, hydrochloric acid, hexane, and dichloromethane were purchased from commercial sources (TCI chemicals, Sigma Aldrich, BLD chemicals, and Merck). Unless specified, HPLC-grade solvents were used for synthesis, recrystallization, and self-assembly. Lasers (405 nm and 650 nm), Arduino Nano, Power supply unit, Photodetector, and Resistors were purchased from ecommerce.

## 2. Instrument Setup:

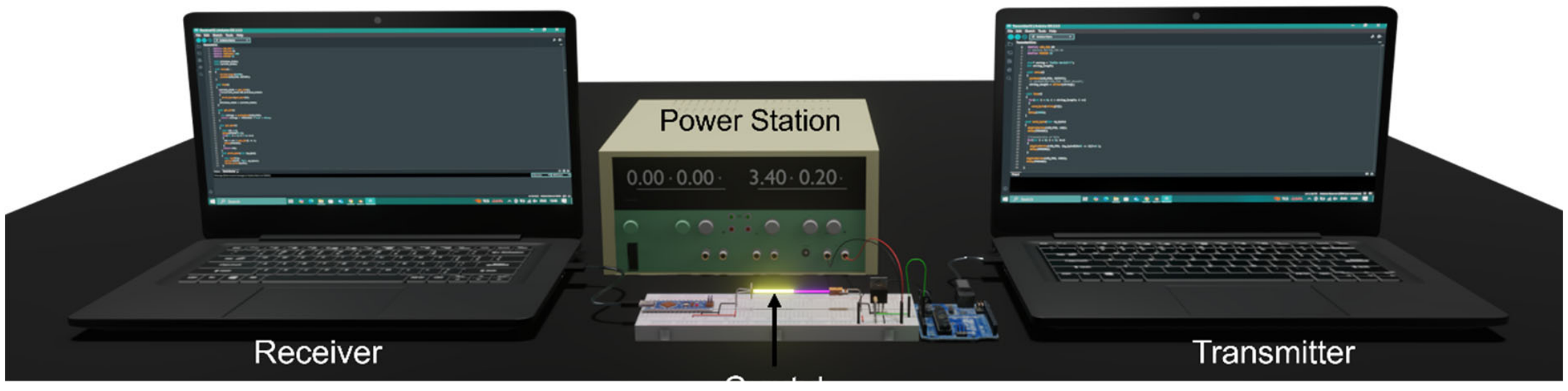

**Scheme S1.** Graphical illustration of the VLC setup

**Preparation of SAA crystals:** For the preparation of SAA crystals, compound 50 mg was dissolved in ethanol: DCM (3:1). The solution was kept in a beaker at ≈8-12°C and left undisturbed for 24 h. Crystal growth was observed, yielding rod-like crystals.

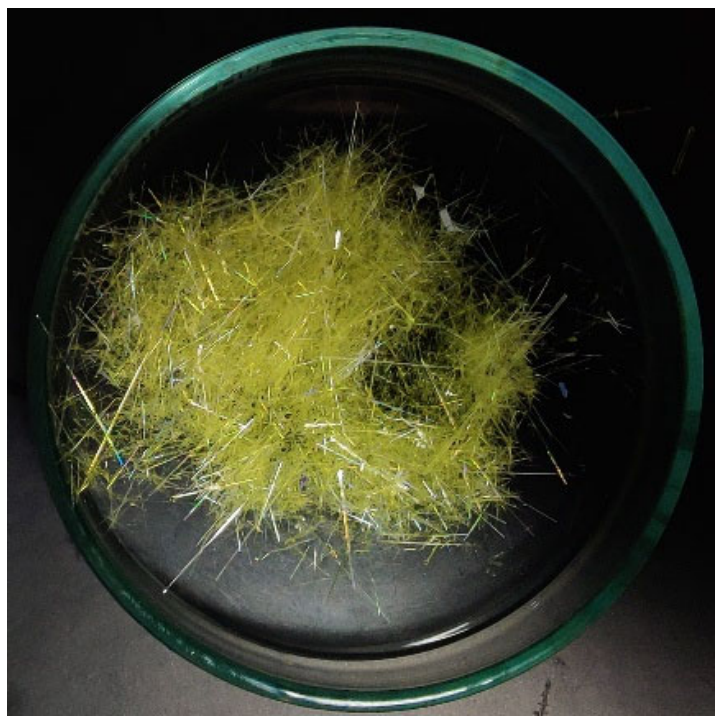

**Figure S1.** Photograph of SAA single crystals

## 3. Optical loss calculations:

The optical loss ($\alpha$) dB of the crystals has been calculated using the equation.

$$10. log_{10}\frac{I_{input}}{I_{output}} = -\alpha L \qquad \text{(Equation 1)}$$

Where the $I_{input}$ and $I_{output}$ are the signal intensities at the input and output point of an SAA crystal waveguide of length (L) of the waveguide.

## 4. Thickness and length-dependent optical loss in SAA organic crystals

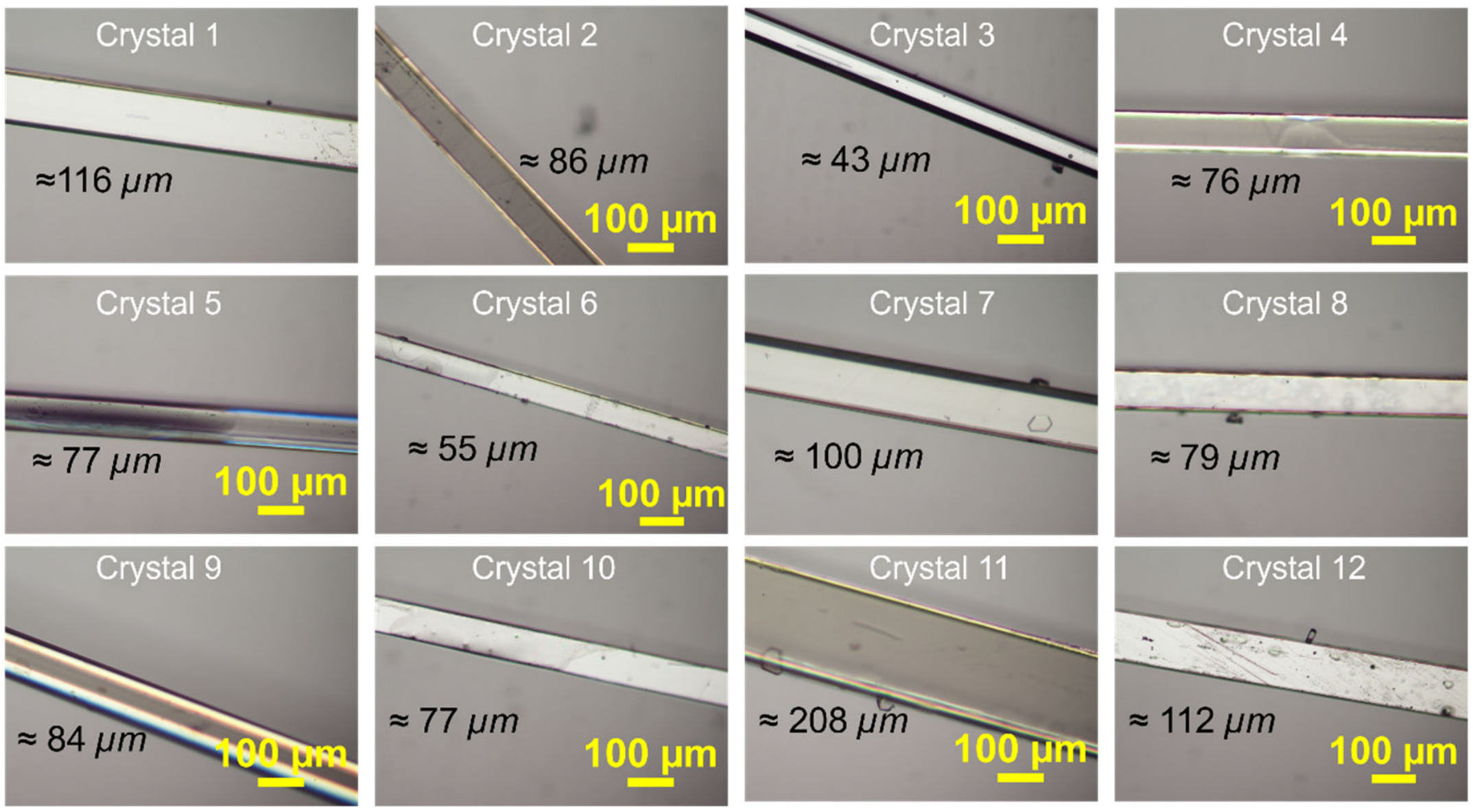

**Figure S2.** Optical images of SAA crystals with varying thickness.

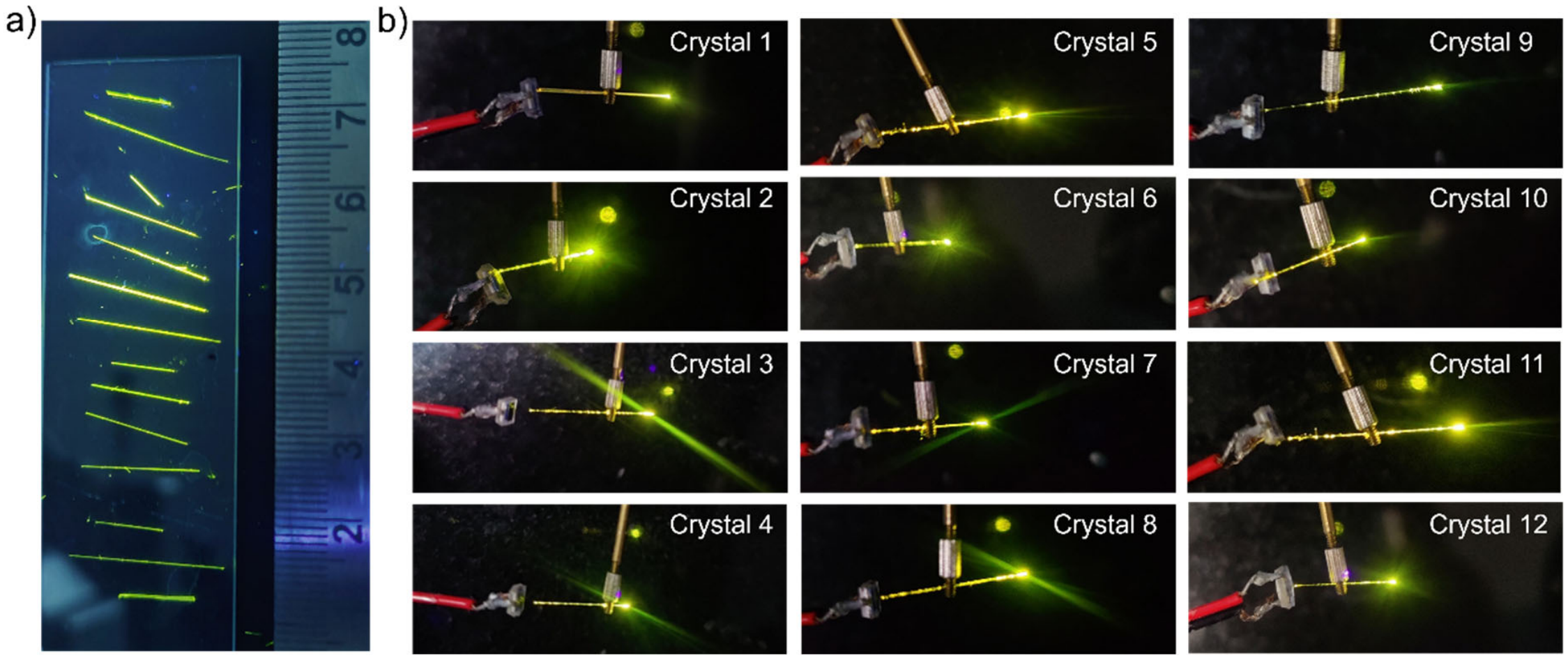

**Figure S3.** a) Photographs of SAA organic single crystals. b) Active waveguiding experiments using crystals of different lengths.

## 5. Angle-dependent optical waveguiding and optical loss of bent SAA crystal

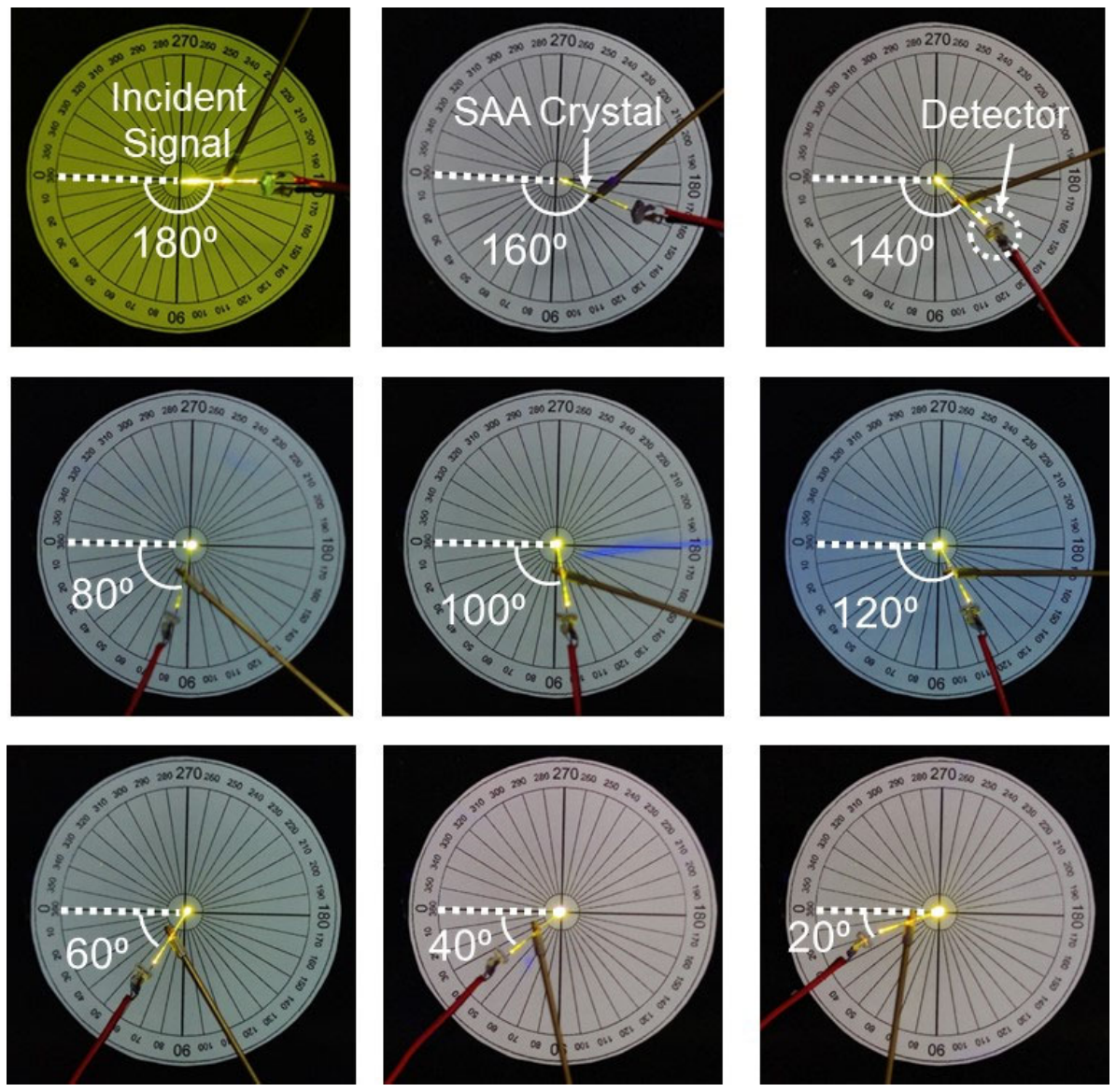

**Figure S4.** Sequential photographs showing the angle between the incident 405 nm light and the SAA single crystal.

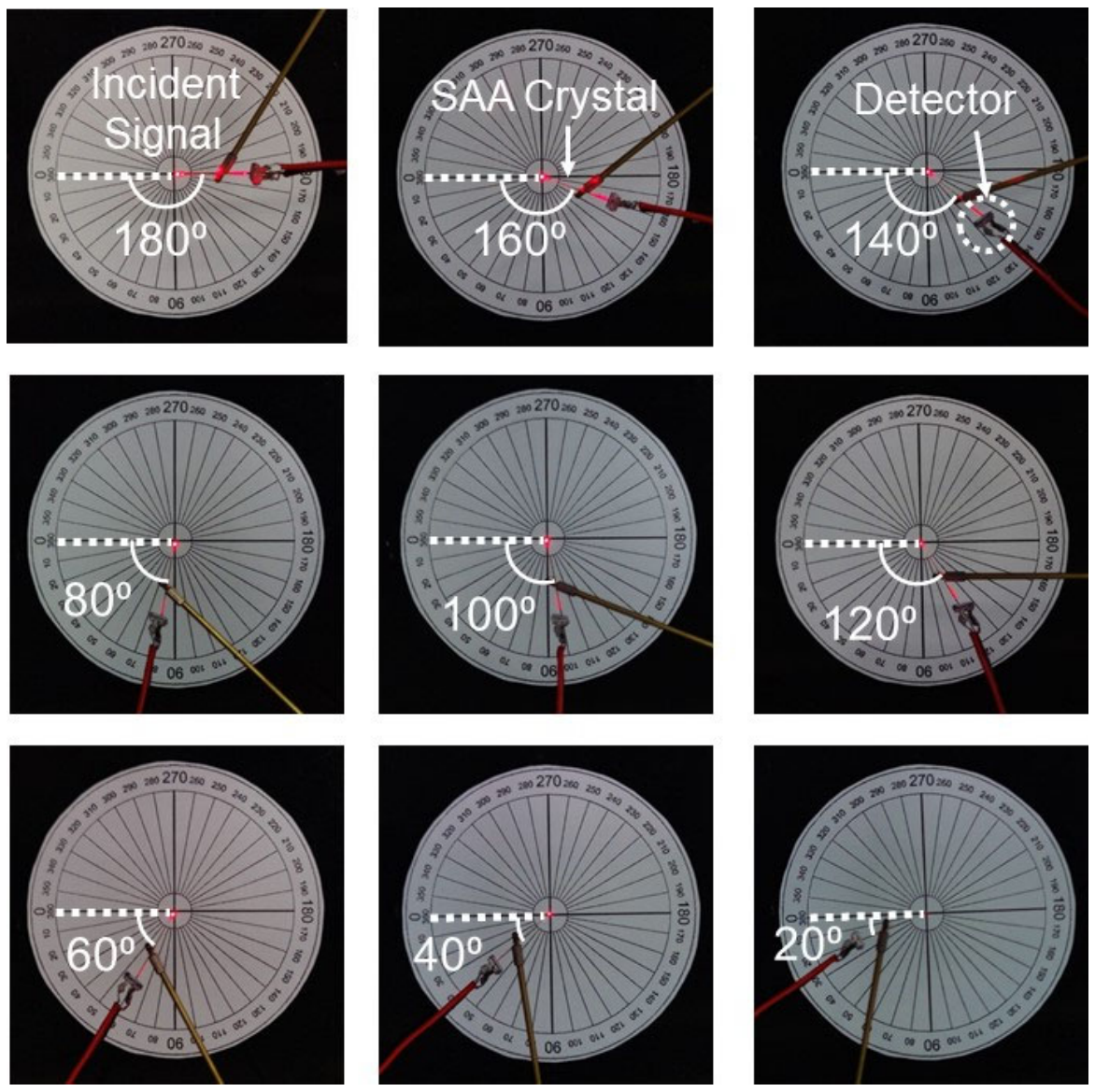

**Figure S5.** Sequential photographs showing the angle between the incident 650 nm light and the SAA single crystal.

## 7. Laser specification

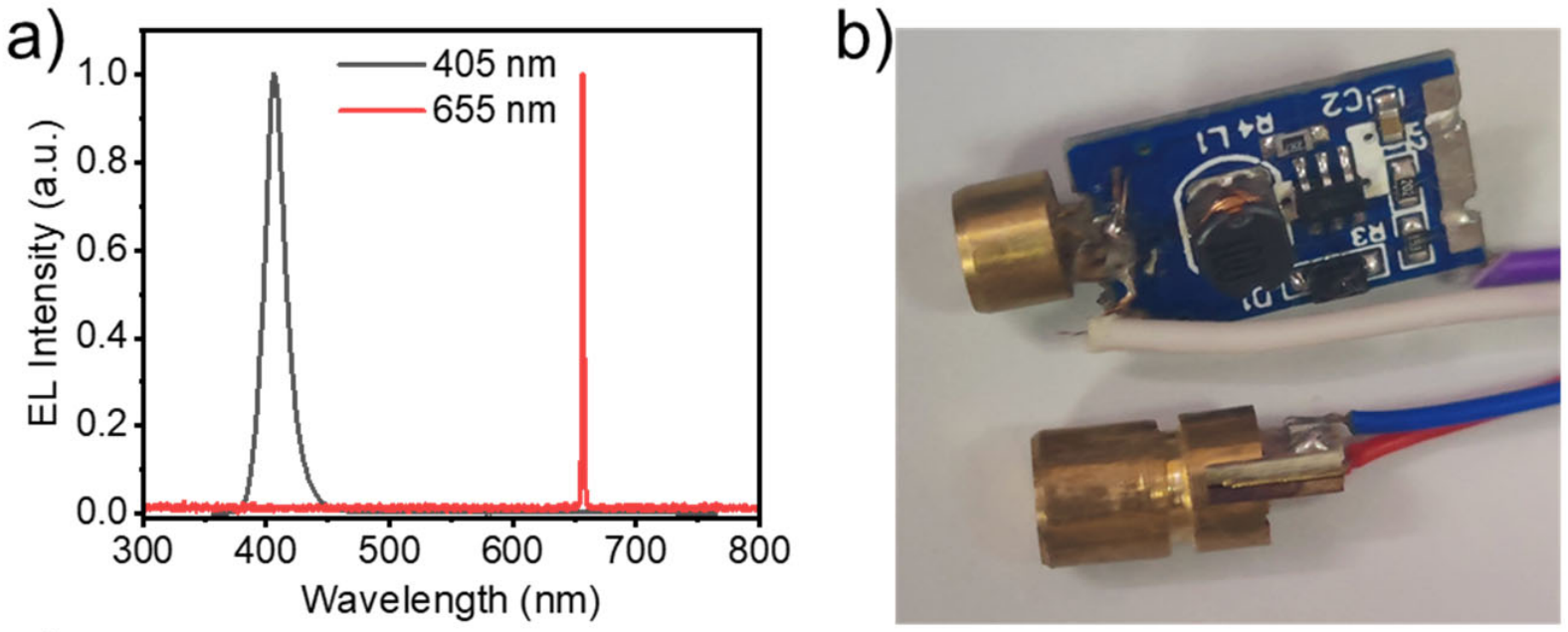

c)

| Electrical/ Optical Description of LED | Symbol | Value | | Unit |
|---|---|---|---|---|
| | | 405 nm | 650 nm | |
| Threshold Voltage | $V_t$ | 3.0 | 5.0 | V |
| Maximum Current | $I_{max}$ | 0.1 | 0.04 | A |
| Wavelength maximum | $\lambda_{max}$ | 405 | 655 | nm |
| FWHM | $\Delta\lambda$ | 20 | 1 | |
| Output Power | P | 50 | 300 | μW |

**Figure S6.** a) Laser EL spectra. b) Photograph of laser module. c) The table shows the electrical and optical specifications of the laser.